\documentclass[apl,reprint,showpacs,twocolumn,superscriptaddress,floatfix]{revtex4-1}
\usepackage{epsfig}
\usepackage{amsmath}
\usepackage{amssymb}
\usepackage{amsfonts}
\usepackage{mathptmx}
\usepackage{dcolumn}
\usepackage{eucal}
\usepackage{bm}
\usepackage{color}
\usepackage[colorlinks,linkcolor=blue,citecolor=blue]{hyperref}

\usepackage{epstopdf}

\usepackage{graphicx}

\usepackage{natbib}

\newcommand{\tmpnote}[1]%
   {\begingroup{\color{blue}\it (FIXME: #1)}\endgroup}

\begin{document}
\title{ Very large thermophase in ferromagnetic Josephson junctions}
\author{F. Giazotto}
\email{giazotto@sns.it}
\affiliation{NEST, Instituto Nanoscienze-CNR and Scuola Normale Superiore, I-56127 Pisa, Italy}
\author{T. T.~Heikkil\"a}
\email{Tero.T.Heikkila@jyu.fi}
\affiliation{Department of Physics and Nanoscience Center, P.O. Box 35
  (YFL), FI-40014 University of Jyv\"askyl\"a, Finland}
\affiliation{Low Temperature Laboratory, Aalto University, P.O. Box 15100, FI-00076 AALTO, Finland}
\author{F. S. Bergeret}
\email{sebastian\_bergeret@ehu.es}
\affiliation{Centro de F\'{i}sica de Materiales (CFM-MPC), Centro
Mixto CSIC-UPV/EHU, Manuel de Lardizabal 4, E-20018 San
Sebasti\'{a}n, Spain}
\affiliation{Donostia International Physics Center (DIPC), Manuel
de Lardizabal 5, E-20018 San Sebasti\'{a}n, Spain}

\begin{abstract}
The concept of \emph{thermophase} refers to the appearance of a phase gradient inside a superconductor originating from the presence of an applied temperature bias across it.  The resulting supercurrent flow may, in  suitable conditions, fully counterbalance the temperature-bias-induced quasiparticle current therefore preventing the formation of any voltage drop, i.e., a thermovoltage, across the superconductor. Yet, the appearance of a thermophase is expected to occur in Josephson-coupled superconductors as well. 
Here we theoretically investigate the thermoelectric response of a thermally-biased Josephson junction based on a ferromagnetic insulator. 
In particular, we predict the occurrence of a \emph{very large} thermophase which can reach  $\pi/2$ across the contact for suitable temperatures and structure parameters, i.e., the
quasiparticle thermal current can reach the critical current. Such a thermophase can be \emph{several} orders of magnitude larger than that predicted to occur in conventional Josephson tunnel junctions.
In order to assess experimentally the predicted very large thermophase we propose a realistic setup realizable with state-of-the-art nano fabrication techniques and well-established materials which is based on a superconducting quantum interference device. This effect could be of strong relevance in several low-temperature applications, for example, for revealing tiny temperature differences generated by coupling the electromagnetic radiation to one of the superconductors forming the junction.

\end{abstract}

\maketitle

Thermoelectric currents in superconductors are often shorted by
supercurrents which generate a phase gradient, a \emph{thermophase}, inside the
superconductor. As suggested a long time ago by Ginzburg \cite{Ginzburg1944}, a
bimetallic superconducting loop constrains the possible phase
gradients, and allows observation of thermoelectric effects via
magnetic fields arising from circulating currents manipulated by
temperature differences. Measurements of such circulating currents
\cite{vanharlingen1980} were larger than that
predicted by theory by several orders of magnitude, a discrepancy that
is yet to be explained \cite{ginzburg2004,galperin2002}. 

In this Letter we
propose an alternative way to produce a very large thermophase in a
SQUID loop consisting of a conventional superconductor in a
spin-polarized contact with a superconductor-ferromagnet bilayer. The
resulting thermophase can reach $\pi/2$ across the contact, i.e., the quasiparticle thermal current can reach the critical current, and can be	
several orders of magnitude larger than in conventional
 Josephson junctions \cite{Guttman,smith1980}. Such a very large thermophase could be used for
detecting tiny temperature differences, for instance, generated by
radiation coupling to one of the superconductors.
\begin{figure}[t!]
\begin{center}
\includegraphics[width=\columnwidth]{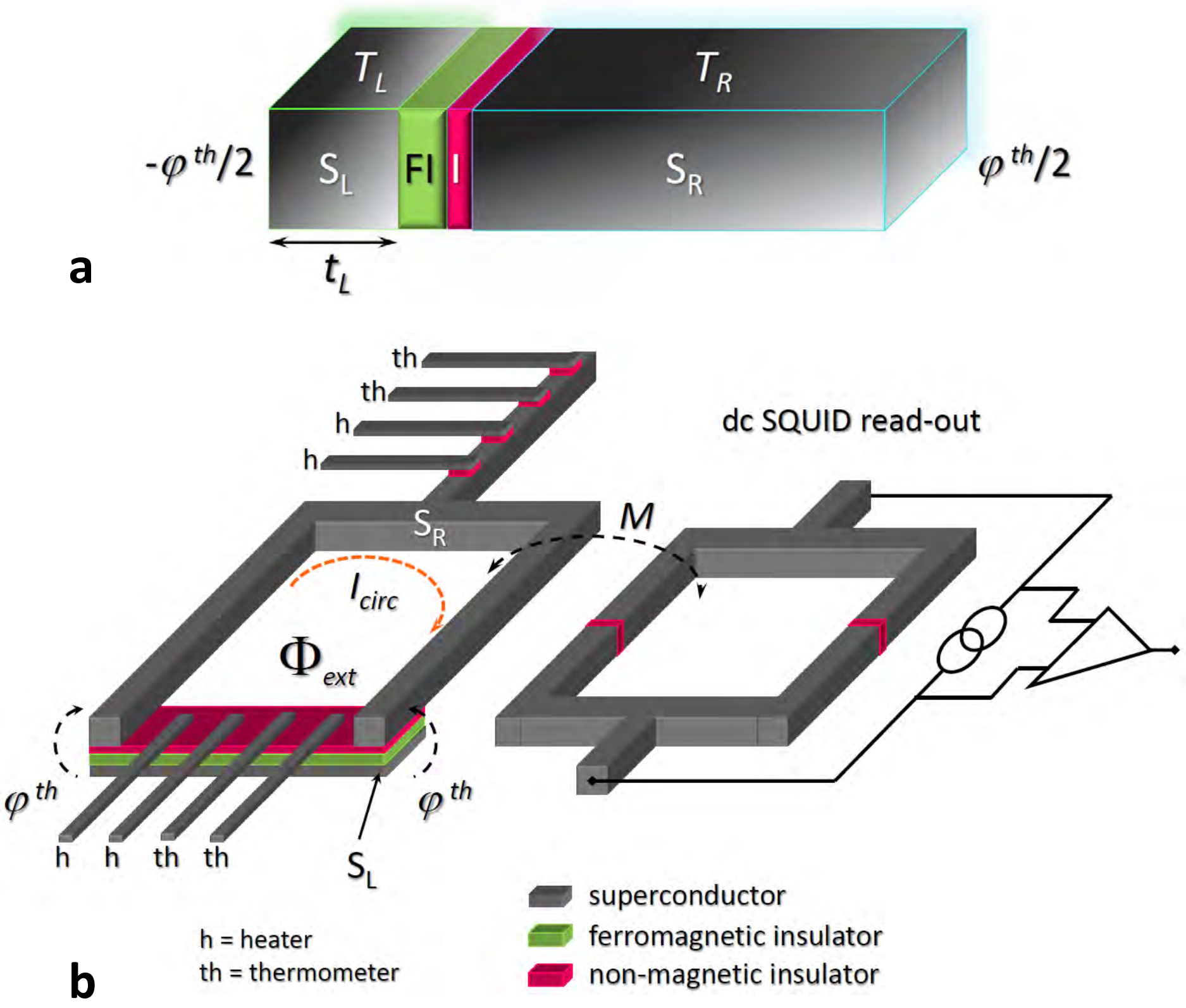}\vspace{-0mm}
\caption{(color online) Thermally-biased ferromagnetic Josephson junction  and the proposed experimental setup. 
(a) Sketch of a generic S-FI-I-S Josephson  junction discussed in the text. It consists of two identical superconductors, S$_\text{L}$ and S$_\text{R}$, tunnel-coupled by a ferromagnetic insulator FI and a non-magnetic barrier I.  The direct contact between FI and   S$_{\text{L}}$ leads to an induced exchange field in the latter, while the non-magnetic  barrier prevents such a field to appear in  S$_{\text{R}}$. 
$T_L$ and $T_R$ are the temperature in S$_{\text{L}}$ and S$_{\text{R}}$, respectively, whereas $\varphi^{th}$ denotes the  \emph{thermophase} originated from thermally biasing the Josephson junction. $t_{{L}}$ is the thickness of S$_{\text{L}}$. 
(b) Scheme of a  detection setup consisting of a temperature-biased superconducting quantum interference device (SQUID) based on the previous junction. 
Superconducting electrodes tunnel-coupled to S$_{\text{L}}$ and S$_{\text{R}}$
serve either as heaters (h) or thermometers (th),
and allow one to impose and detect a temperature gradient across the SQUID.
The magnitude of the induced $\varphi^{th}$ can be determined
 by measuring variations of the supercurrent ($I_{circ}$) circulating in the interferometer through a conventional dc SQUID
 inductively coupled to the first ring.
 $M$ denotes the mutual inductance between the loops, and $\Phi_{ext}$ is the external applied magnetic flux.\label{fig1}}
\end{center}
\end{figure}

We consider a Josephson junction [see Fig.~\ref{fig1}(a)] consisting of two  superconductors S$_\text{L}$ and S$_\text{R}$ tunnel coupled  through  a ferromagnetic insulator (FI) and a non-magnetic (I)  barrier.  
The FI  has different transmissivities for spin-up and spin down electrons and therefore acts as a spin-filter.\cite{MooderaRev2007}  
The interaction between the conduction electrons in  S$_\text{L}$  with the localized magnetic moments of the FI leads to an effective spin-splitting field $h$ in the left electrode that decays away from the interface  over the superconducting coherence length $\xi_0$.\cite{Tokuyasu1988} 
 The thin I layer placed on the right side of the FI prevents such a spin-splitting field to be induced in S$_\text{R}$\cite{Moodera1990,Moodera2013,BinLi2013}.
We assume that the thickness $t_L$ of S$_{\text{L}}$ is smaller than $\xi_0$ so that the induced $h$ is spatially uniform  across the entire  S$_{\text{L}}$ layer \cite{Giazotto2008}. 
The junction is temperature biased, so that $T_{L,R}$ is the temperature
in S$_{\text{L,R}}$, respectively, and $\varphi^{th}$ denotes the phase difference
between the superconducting order parameters induced by such a
temperature difference.
We focus on the \emph{static} (i.e., time-independent) regime so that a dc Josepson current can flow in response to the applied thermal gradient but no thermovoltage  develops across the junction.
   
In order to analyze the setup, we generalize the calculation of
\cite{Ozaeta2014} to the case of two superconductors. The  total electric current $I$ flowing through the junction is given by  the sum of the quasiparticle ($I_{qp}$)  and the Josephson contribution ($I_J$)
\begin{equation}
I=I_{qp}+I_{J}=I_{qp}+I_c\sin{\varphi^{th}}, 
\label{Itot}
\end{equation}
where $I_c$ is the critical  supercurrent.
The current contribution  proportional to
$\cos{\varphi^{th}}$\cite{BaroneBook} does not contribute, since it does not  possess any
thermoelectric response, and it would require a
finite voltage.
The explicit forms for $I_{qp}$ and $I_c$ in Eq. (\ref{Itot}) can be
obtained from the expressions for the current through a spin-filter
barrier with polarization $P$
\cite{BVV2012,BVV2012bis,Machon2013},
\begin{equation}
I_{qp}=\frac{P}{2eR_T}\int^{\infty}_{-\infty} d\varepsilon N_{L}^{-}(\varepsilon)N_R(\varepsilon)\left[\textbf{f}_L(\varepsilon)-\textbf{f}_R(\varepsilon)\right]
\label{qpcurrent}
\end{equation}
and
\begin{eqnarray}
I_{c}=-\frac{\sqrt{1-P^2}}{2eR_T}\int^{\infty}_{-\infty} d\varepsilon[\text{Re}\,M_{L}(\varepsilon)\text{Im}\,F_{R}(\varepsilon)\textbf{f}_L(\varepsilon)\nonumber\\
+\text{Im}\,M_{L}(\varepsilon)\text{Re}\,F_{R}(\varepsilon)\textbf{f}_R(\varepsilon)].
\label{Jcurrent}
\end{eqnarray}
In Eqs. (\ref{qpcurrent}) and (\ref{Jcurrent}), 
$N_{L}^{-}(\varepsilon)=\left[N_L(\varepsilon+h)-N_L(\varepsilon-h)\right]/2$, $N_{L,R}(\varepsilon)=|\text{Re}[(\varepsilon+i\Gamma)/\sqrt{(\varepsilon+i\Gamma)^2-\Delta^2_{L,R}}]|$ is the normalized BCS  density of states in S$_\text{L,R}$, $M_L(\varepsilon)=[F_L(\varepsilon+h)+F_L(\varepsilon-h)]/2$,
 $F_{L,R}(\varepsilon)=\Delta_{L,R}/\sqrt{(\varepsilon+i\Gamma)^2-\Delta^2_{L,R}}$,
  $\textbf{f}_{L,R}(\varepsilon)=\tanh(\varepsilon/2k_BT_{L,R})$, and
  $\Delta_{L,R}$ is the energy gap in S$_\text{L,R}$ which has to be
  determined self-consistently due to the presence of $h$ and finite $T_{L/R}$. For ideal superconductors $\Gamma
  \rightarrow 0^+$ \cite{Dynes1984}; we have checked that the value
  $\Gamma=10^{-4}\Delta_0$ chosen in our numerics  negligibly
  affects the results.
  Above, $\Delta_0$ denotes the zero-temperature, zero-exchange field superconducting energy gap.
Furthermore, $e$ is the electron charge, $k_B$ is the Boltzmann constant, and $R_T$ is the normal-state resistance of the junction. 
Equation (\ref{qpcurrent}) shows that for a non-vanishing $I_{qp}$ to exist  (i) a finite $h$ should be induced in one of the superconductors, and (ii)  $P$ has to be finite.\cite{Ozaeta2014}  

In an \emph{electrically-open} configuration the total charge current has to
vanish, $I=0$. Therefore, in order to ensure  a vanishing thermovoltage across the
junction, the  quasiparticle current $I_{qp}$ induced by the temperature gradient  has to be  canceled  by an opposite dc supercurrent $I_J$. This  cancelation  is the origin of the thermophase, which is defined as
\begin{equation}
\varphi^{th}=\arcsin\left(-\frac{I_{qp}}{I_c}\right).
\label{thermophase}
\end{equation}
This thermophase is thus a measure of the amplitude of the
thermoelectric effect at the contact between the superconductors.
In the \emph{electrically-closed} configuration $\varphi^{th}$ is necessarily no longer the phase difference across the junction. In this case Eq. (\ref{thermophase}) should be viewed as a definition, characterizing the relative magnitude of the quasiparticles current vs. the critical current of the junction. Below we discuss a scheme for measuring this thermophase.

\begin{figure}[t!]
\begin{center}
\includegraphics[width=\columnwidth]{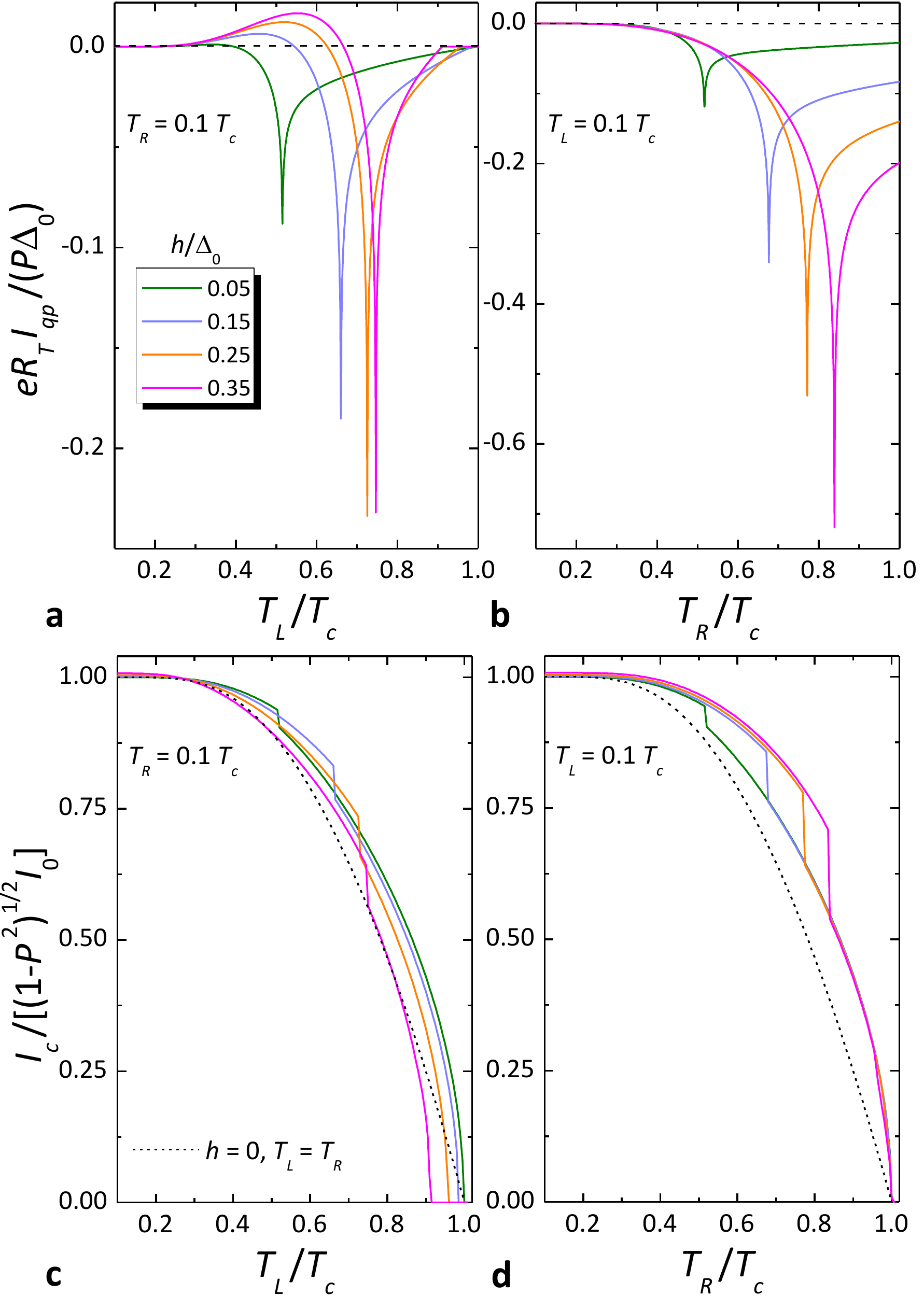}\vspace{-1mm}
\caption{(color online) Quasiparticle and Josephson critical currents under thermal-bias conditions for different values of the exchange field. 
Quasiparticle current $I_{qp}$ 
(a) vs $T_L$  at $T_R=0.1 T_c$
and 
(b) vs $T_R$ at $T_L=0.1 T_c$; Josephson critical current
$I_c$ (c) vs $T_L$  at $T_R=0.1 T_c$ and 
(d) vs $T_R$  at $T_L=0.1 T_c$. 
Dash-dotted curves in panels (c) and (d) are calculated for $h=0$ and $T_L=T_R$. 
$\Delta_0$ denotes the zero-exchange field, zero-temperature
superconducting energy gap corresponding to the critical temperature
$T_c \approx \Delta_0/1.764k_B$ whereas $I_0$ is the zero-exchange field, zero-temperature Josephson critical current. 
}
\label{fig2}
\end{center}
\end{figure}

Let us analyze  the behavior of $I_{qp}$ and of $I_c$ under thermal bias.
Figures \ref{fig2}(a) and \ref{fig2}(b)  show $I_{qp}$ vs  $T_L$ and
$T_R$, respectively,  when the temperature of  the other electrode is
fixed to $0.1 T_c$. 
Here $T_c$ is the superconducting critical temperature in the absence of $h$ which we assume, for simplicity, to be the same for both superconductors. 
By varying $T_L$, and depending on the temperature range, the thermocurrent can be either \emph{positive}, i.e., flowing according to the thermal gradient set across the junction, or \emph{negative}. The sign of the current can be ascribed to a more \emph{electron}- or \emph{hole}-like contribution to thermoelectric transport, respectively. 
The amplitude of $I_{qp}$ is, in general, larger for larger $h$, and drops 
 eventually to zero at the  temperature for which  $\Delta_L$
 vanishes. The value of such a critical temperature depends on the value of $h$. 
By contrast,  Fig. \ref{fig2}(b) shows that   the thermocurrent  does
not change sign if $T_L$ is  held constant and $T_R(>T_L)$ is varied. The  amplitude of $I_{qp}$  in this case is  larger than that obtained by varying $T_L>T_R$,  
and is finite  at $T_c$. 
Notably, the quasiparticle characteristics exhibits sharp dips positioned at the temperatures satisfying  the condition 
\begin{equation}
|\Delta_L(T_L,h)-\Delta_R(T_R)|=h\;.
\label{condition}
\end{equation} 
 The large values obtained by  $I_{qp}$ are  the origin of a \emph{very large} thermophase achievable in ferromagnetic Josephson junctions. 
\begin{figure}[t!]
\includegraphics[width=\columnwidth]{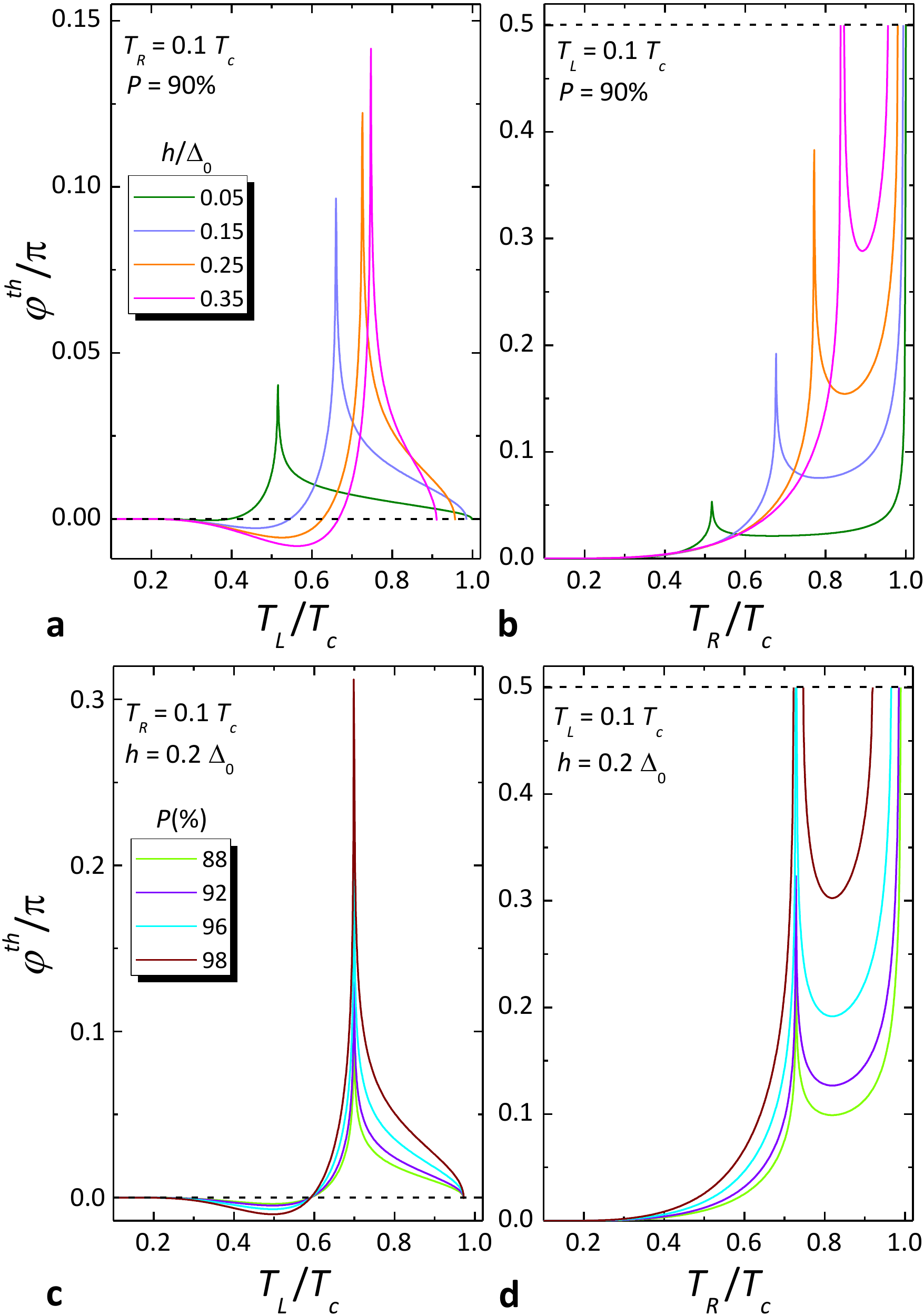}\vspace{-1mm}
\caption{(color online) Behavior of the thermophase in a ferromagnetic Josephson junction. 
(a) Thermophase $\varphi^{th}$  vs $T_L$ calculated for several values of the exchange field $h$ at $T_R=0.1 T_c$ and $P=90\%$.
(b) $\varphi^{th}$  vs $T_R$ calculated at $T_L=0.1 T_c$ and $P=90\%$ for the same exchange field values as in panel (a).
(c) $\varphi^{th}$  vs $T_L$ calculated for several values of the  polarization  $P$ of the spin-filter barrier at $T_R=0.1 T_c$ and $h=0.2\Delta_0$.
(d) $\varphi^{th}$  vs $T_R$ calculated at $T_L=0.1 T_c$ and $h=0.2\Delta_0$ for the same $P$ values as in panel (c). $\varphi^{th}$ is not defined in the temperature regions where $|I_{qp}|$ exceeds $I_c$.
}
\label{fig3}
\end{figure}

Similarly, the lower panels of Fig. \ref{fig2} display the behavior of $I_c$ vs  $T_R$ and $T_L$ by holding the other electrode  at $0.1T_c$. 
The $I_c(T)$ curves differ drastically from those obtained for $h=0$ at $T_L=T_R$ (dash-dotted curves). 
In particular, at a low enough temperature  $I_c$ gets larger by increasing $h$. 
This remarkable effect corresponds to the supercurrent enhancement discussed in Ref. \cite{splitsupercurrent} which occurs even 
 for $P=0$. 
We stress that the $I_c$ strenghtening  joined with the sharp jumps
appearing at those temperatures where Eq.~(\ref{condition}) holds are a manifestation  of an exchange field induced in S$_{\text{L}}$  and of a nonequilibrium condition  stemming from the thermal bias \cite{splitsupercurrent}.

The thermophase $\varphi^{th}$ is obtained from Eq. (\ref{thermophase}). 
 Figure \ref{fig3}(a) and (b) shows the dependence of $\varphi^{th}$   on $T_L$ and $T_R$, respectively, when the other electrode is kept  at 
$0.1T_c$. We have chosen a reasonable  polarization of the barrier
($P=90\%$) and moderate $h$ values easily achievable with present-day
experiments. \cite{MooderaRev2007,Moodera2013} 
We find that $\varphi^{th}$ can be vary large, close to $\pi/2$ for $h\gtrsim 0.3\Delta_0$.  
This substantial effect has to be compared to the minute one achievable in conventional non-ferromagnetic Josephson tunnel junctions where  $\varphi^{th}\sim 10^{-4}$  is expected.\cite{Guttman,smith1980} 
We also notice the presence of temperature regions where
$\varphi^{th}$ is not defined since $I_{qp}$ may exceed $I_c$ [see
Fig.~\ref{fig3}(b)]. In this case a finite dc voltage is induced across
the contact.

\begin{figure}[t!]
\includegraphics[width=\columnwidth]{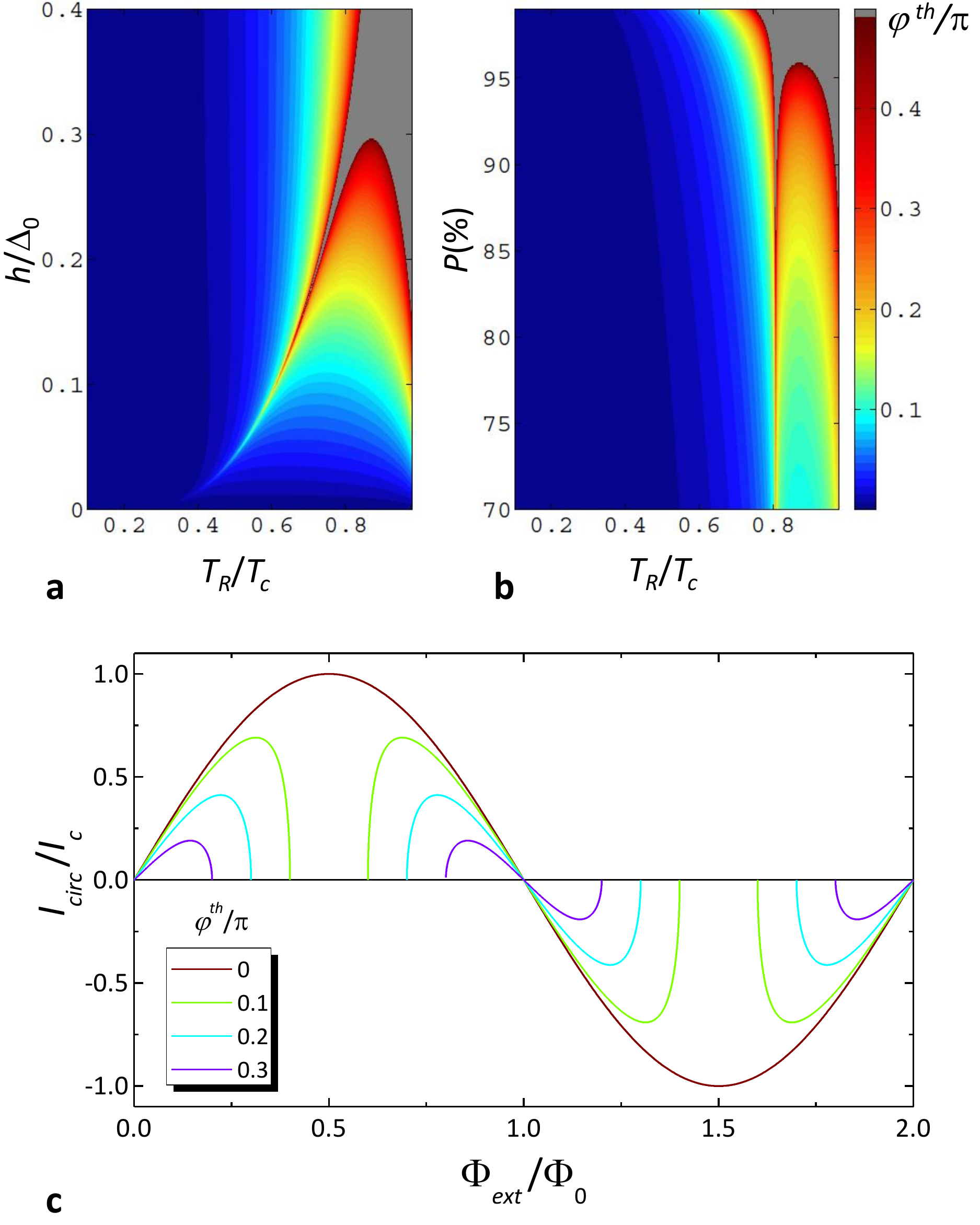}\vspace{-1mm}
\caption{(color online) Full behavior of the thermophase and response of the circulating current in a temperature-biased FI SQUID. 
(a) Color plot of the thermophase $\varphi^{th}$  vs $T_R$ and $h$ calculated at $T_L=0.1 T_c$ for $P=96\%$.
(b) Color plot of $\varphi^{th}$  vs $T_R$ and $P$ calculated at $T_L=0.1 T_c$ for $h=0.3\Delta_0$.
(c) Circulating current $I_{circ}$ vs external magnetic flux $\Phi_{ext}$ for a few values of the
thermophase in a symmetric SQUID.
}
\label{fig4}
\end{figure}

The impact of $P$ on $\varphi^{th}$ is shown in panel (c) and (d) of Fig. \ref{fig3} where we set $h=0.2\Delta_0$, and the temperature in the superconductors is varied similarly to panel (a) and (b), respectively. 
The increase of  $P$ leads to a sizable thermophase enhancement, as $I_{qp}/I_c\propto P/\sqrt{1-P^2}$ [see Eqs. (\ref{qpcurrent}-\ref{thermophase})].  
From Fig.~\ref{fig3} it becomes clear that large values of  $\varphi^{th}$ can  be obtained  more easily  by increasing $T_R$ 
 while keeping S$_{\text{L}}$ at a low temperature, consistently with the   $I_{qp}(T_R)$ dependence shown in  Fig.~\ref{fig2}(b). 
The full behavior of $\varphi^{th}$ vs $T_R$ for  $T_L=0.1T_c$ is displayed in the color plots of Fig. \ref{fig4}. 
Specifically,  we set $P=96\%$ and varied $h$ in panel (a), whereas in panel (b) we set $h=0.3\Delta_0$ and varied $P$.
The figures show that a sizeable $\varphi^{th}$ can be obtained in a rather large range of  parameters,
 and may provide a valuable tool for tailoring optimized junctions where $\varphi^{th}$ is maximized.

To assess experimentally the predicted very large thermophase we propose
 the setup depicted in Fig.~\ref{fig1}(b). 
It consists of a superconducting quantum interference device (SQUID)
including a FI, and comprising a number of superconducting  tunnel junctions which 
can either heat or perform accurate electron thermometry \cite{Giazotto2006}. 
From the materials side, FIs such as  EuO or EuS (providing $P$ up to
$\sim 98\%$) \cite{Santos2008} in contact with superconducting Al appear as
ideal candidates for the implementation of the structure which can be
realized with standard lithographic techniques. 
The ratio $h/\Delta_0$ in such structures depends on the thickness 
of the Al layer and quality of the contact.
In the superconducting state of Al, values ranging from $h/\Delta_0\approx 0.2$ up to $0.6$
have been reported. \cite{Catelani2011,Adams2013,Moodera2013,BinLi2013}
Alternatively,  GdN
barriers  in combination with Nb or NbN  could be used as
well.\cite{Blamire2011,Blamire2014} 

In the SQUID, the thermophase $\varphi^{th}_{1/2}$ developed across
the two junctions results into a nonzero total circulating current $I_{circ}$. 
In the absence of an external flux and for negligible loop inductance, the amplitude of the total circulating current
is given by
\begin{equation}
I_{circ}=\frac{I_{c1}I_{c2}}{I_{c1}+I_{c2}} |\sin(\varphi_1^{
  th})-\sin(\varphi_2^{th})|,
\end{equation}
where $I_{ci}$ is the critical current for contact $i=1,2$. Since $\varphi^{ th}_i$ depends on
the ratio of the quasiparticle and critical supercurrents, an
asymmetry of the resistances between the contacts would not cause a circulating
current. However, replacing one of the junctions, say 2, by a conventional SIS
junction would set $\varphi_{2}^{th} \approx 0$ and therefore would lead to
a large $I_{\rm circ}$ even without an external flux.

An alternative way to measure the thermophase is to consider the case of a
finite external flux $\Phi_{ext}$. If the SQUID junctions are identical,  $I_{circ}$ can be
written as
\begin{equation}
I_{circ}=I_c(\sin\varphi_1-\sin\varphi^{th})=-I_c(\sin\varphi_2-\sin\varphi^{th}).
\label{Icircflux}
\end{equation}
In Eq. (\ref{Icircflux}), $I_{circ}$ is written as the sum of a magnetic flux-dependent part and the one of thermoelectric origin,
and the second equality expresses the conservation of the supercurrent circulating along the loop through each junction of the SQUID. 
In the limit of negligible ring inductance Eq. (\ref{Icircflux}) can ben analytically solved for $I_{circ}$
 by imposing  fluxoid quantization,  $\varphi_2-\varphi_1=2\pi
\Phi_{ext}/\Phi_0$, where $\Phi_0=2.067\times 10^{-15}$ Wb is the flux quantum. 
The result for $I_{circ}$ is thus
\begin{equation}
I_{circ}=I_c\sin \left(\frac{\pi \Phi_{ext}}{\Phi_0}\right)\text{Re}\left[\sqrt{1-\frac{\sin ^2\varphi^{th}}{\cos^2\left(\frac{\pi \Phi_{ext}}{\Phi_0}\right)}}\right ],
\label{circulating}
\end{equation}
which holds for $I_{qp}\leq I_c$ [see Eq.~(\ref{thermophase})]. 
For $\varphi^{th}=0$
Eq. (\ref{circulating}) reduces to 
\begin{equation}
I_{circ}=I_c\sin \left(\frac{\pi\Phi_{ext}}{\Phi_0}\right ). 
\end{equation}
As shown in Fig.~4(c), the
presence of a finite thermophase gives rise to regions of flux close
to $\Phi_0/2$ and $3\Phi_0/2$ where $I_{circ}$
\emph{vanishes} because the thermoelectric current becomes larger than the
effective critical current of the SQUID. The presence of these regions is a direct evidence of the thermophase.
The circulating current can be detected  through a conventional dc SQUID inductively-coupled to the first loop  [see Fig. \ref{fig1}(b)] so that
 $\Phi_{SQUID}\sim M I_{circ}$, where $\Phi_{SQUID}$ is the magnetic flux induced in the read-out SQUID and $M$ is the mutual inductance coefficient. 
For instance, for typical values of  $I_c=1\,\mu$A and  $M=10$ pH, $\Phi_{SQUID}$ up to $\sim 10^{-17}\,\text{Wb}\approx 5\times 10^{-3}\Phi_0$ can be generated with a proper temperature bias.  
This can be well detected with standard SQUIDs which provide routinely magnetic flux sensitivities $\sim 10^{-6}\Phi_0/\sqrt{\text{Hz}}$ \cite{clarkbook}. 

In conclusion, we have predicted the occurrence of a \emph{very large} thermophase in thermally-biased Josephson junctions based on FIs. 
This sizeable effect can be detected in a structure realizable with current state-of-the-art nanofabrication techniques and well-established materials. 
Besides shedding  light onto fundamental problems related to the
\emph{thermoelectric} response of superconductors and exotic
weak links in the Josephson regime, the very sharp thermophase
response (see Fig.~3) combined with the low heat capacity of
superconductors could allow realizing ultrasensitive radiation
detectors \cite{Giazotto2006} where radiation induced heating of one of the
superconductors is detected via the thermophase. 
The presence of magnetic
material also allows for adding a new control parameter to the
experimental investigation of coherent manipulation of heat flow at the nanoscale \cite{Giazotto2012,Martinez2014,Giazotto2014,Meschke2006}.

F.G. acknowledges the Marie Curie Initial Training Action (ITN) Q-NET
264034, and the European Research Council under the European Union's Seventh Framework Programme (FP7/2007-2013)/ERC grant agreement No. 615187-COMANCHE for partial financial support. The work of F.S.B has been
supported by the Spanish Ministry of Economy and Competitiveness under
Project No. FIS2011-28851-C02-02 and the work of T.T.H. by the ERC
(Grant No. 240362-Heattronics) and the Academy of Finland through its
Center of Excellence program.

\end{document}